\documentclass[aps,showpacs,epsfig]{revtex4}
\usepackage{epsfig}

\newcommand{\be}{\begin{equation}}
\newcommand{\ee}{\end{equation}}
\newcommand{\bea}{\begin{eqnarray}}
\newcommand{\eea}{\end{eqnarray}}
\usepackage{color}
\begin{document}

\title{\bf\Large {Quantum critical behavior  in three-dimensional one-band Hubbard model at half-filling}
}

\author{Naoum Karchev\footnote{Tel: +3 592 8527754 \\ E-mail address: naoum@phys.uni-sofia.bg } }

\affiliation{ Department of Physics, University of Sofia, 1126 Sofia, Bulgaria }

\begin{abstract}
A one-band Hubbard model with hopping parameter $t$ and Coulomb repulsion $U$ is considered at half-filling. By means of the Schwinger bosons  and slave fermions representation of the electron operators and integrating out the spin-singlet Fermi fields an effective Heisenberg model with antiferromagnetic exchange constant is obtained for vectors which identifies the local orientation of the spin of the itinerant electrons. The amplitude of the spin vectors is an effective spin of the itinerant electrons accounting for the fact that some sites, in the ground state, are doubly occupied or empty. Accounting adequately for the magnon-magnon interaction the N\'{e}el temperature is calculated. When the ratio $\frac tU$ is small enough ($\frac tU\leq 0.09$) the effective model describes a system of localized electrons. Increasing the ratio increases the density of doubly occupied states which in turn decreases the effective spin and N\'{e}el temperature. The phase diagram in the plane of temperature $\frac {T_N}{U}$ and parameter $\frac tU$ is presented. The quantum critical point ($T_N=0$) is reached at $\frac tU=0.9$. The magnons in the paramagnetic phase are studied and the contribution of the magnons' fluctuations to the heat capacity is calculated. At the N\'{e}el temperature the heat capacity has a peak which is suppressed when the system approaches a quantum critical point. It is important to stress that, at half-filling, the ground state, determined by fermions, is antiferromagnetic. The magnon fluctuations drive the system to quantum criticality and when the effective spin is critically small these fluctuations suppress the magnetic order.

\end{abstract}

 \maketitle

\section {\bf Introduction}

Quantum phase transitions arise in many-body systems because of competing interactions that support different ground states. At the quantum critical point (QCP)the matter undergoes a transition from one phase to another at zero temperature.
A nonthermal control parameter, such as pressure, drives the system to the QCP. The typical temperature-pressure phase diagrams observed in the heavy-fermion materials $CePd_2Si_2$, $CeIn_3$, $CeRh_2Si_2$, $CeCu_2Si_2$ and $CeRhIn_5$ \cite{Grosche96,Movshovich96,Mathur98,Grosche01,Araki02,Yashima07} show that at ambient pressure the compounds order into antiferromagnets below the N\'{e}el temperature $T_N$. Applying pressure reduces $T_N$ monotonically. The QCP is the critical pressure at which the N\'{e}el temperature $T_N=0$.

The quantum critical behavior has been extensively studied for many years.
Many books \cite{Moriya85,Sachdev99}, review articles \cite{Lonzarich97,Grigera01,Vojta03,GvL07,Pfleiderer09,Si10,Knebel11}  and papers investigate systematically the magnetic quantum critical point.

The magnetism of cerium based compounds is determined by the $4f$ electrons of $Ce^{3+}$ ions. The strong spin-orbit coupling  splits the $4f$ electrons into $j=\frac 52$ and $j=\frac 72$ multiplets , where $j$ is the total angular momentum. Only the sextuplet effectively contributes to the low energy excitations. It is further split into a $\Gamma_7$ doublet and $\Gamma_8$ quadruplet due to the crystal electric field. For isotropic systems like $CeIn_3$, the energy level of $\Gamma_7$ is lowest. The eigenstates are  $|\Gamma_{7\pm}>=\sqrt{\frac 16}|\pm \frac 52>-\sqrt{\frac 56}|\mp \frac 32>$, where "+" and "-" denote up and down "pseudo-spins" respectively. The three-dimensional (3D) one-band Hubbard model is the simplest model of itinerant magnetism of the isotropic cerium based compounds. Although the hybridization with $In5p$ electronic states may be important, here it is considered as a renormalization of the hopping amplitude.

The ground state of the 3D Hubbard model on a simple cubic lattice, at half-filling
has antiferromagnetic long range order for all positive values of the onsite Coulomb repulsion. There have been many attempts to calculate the N\'{e}el temperature $T_N$ using  quantum Monte Carlo(QMC) simulations \cite{Hirsh87,Scalettar89,Ulmke96,Staudt00},
variational methods \cite{Kakehashi85,Kakehashi86}, Hartree Fock theory \cite{Dongen91}, strong coupling expansions
\cite{Szczech95}, and dynamical mean field theory (DMFT) \cite{Jarrell92,Georges93,Ulmke95}. Calculations beyond the dynamical mean field theory-the dynamical cluster approximation (DCA) \cite{Kent05}, cluster generalization of the DMFT \cite{Fuchs11} and dynamical vertex approximation
$(D\Gamma A)$ \cite{Toschi07,Toschi11} have been proposed. All these studies do not show any trace of a quantum critical point.

Here we focus attention on the quantum critical behavior in itinerant antiferromagnets. The increasing of the double occupancy in the $3D$ one-band Hubbard model at half-filling pushes the system to the quantum criticality. Increasing the ratio $\frac tU$, where $t$ is the hopping parameter and $U$ is the Coulomb repulsion, increases the density of doubly occupied states which in turn decreases the effective spin and N\'{e}el temperature. The Quantum Critical Point is a state with a critically high density of the doubly occupied state.

It is important to stress that, at half-filling, the ground state, determined by fermions, is antiferromagnetic. The magnon fluctuations drive the system to quantum criticality and when the effective spin is critically small these fluctuations suppress the magnetic order. Magnon formation and the effects of magnons' fluctuations are non-perturbative phenomena even at small values of $U/t$ and cannot be obtained within perturbation theory in analogy, for example, with the weak coupling
theory of superconductivity. One of the important results in the paper is the explicit account for  magnons' fluctuations.
We employ a technique of calculation, which captures the essentials of the magnons' fluctuations in the theory, and for $2D$ systems one obtains zero N\'{e}el temperature, in accordance with the Mermin-Wagner theorem\cite{M-W}.

The paper is organized as follows. In Sec. II, starting from the one-band Hubbard model at half-filling ,with hopping parameter $t$ and Coulomb repulsion $U$, we derive an effective Heisenberg-like model, with antiferromagnetic exchange constant, in terms of the vector describing the local orientations of the magnetization. The transversal fluctuations of the vector are the magnons in the theory. The amplitude $m$ of the spin vectors is an effective spin of the itinerant electrons accounting for the fact that some sites, in the ground state, are doubly occupied or empty. This is a base for N$\acute{e}$el temperature calculation. Section III is devoted to $T_N/J-m$ and $T_N/U-t/U$ phase diagrams of the model. The paramagnetic phase of itinerant antiferromagnets is explored in Section IV. The calculations of the specific heat for different values of the control parameter $t/U$  are presented in Section V.
A summary in Sec. VI concludes the paper.

\section {\bf Effective model}

We consider a theory with Hamiltonian
\be \label{QCB1}
h  = -t\sum\limits_{\langle ij \rangle} \left( c_{i\sigma }^ + c_{j\sigma}  + h.c. \right)
+ U \sum\limits_i n_{i\uparrow} n_{i\downarrow} -\mu \sum\limits_i n_i\ee
where $c_{i\sigma }^+$ and $c_{i\sigma }$ ($\sigma=\uparrow,\downarrow$) are creation and annihilation operators for spin-1/2 Fermi operators of itinerant electrons, $n_{i\sigma}=c^+_{i\sigma}c_{i\sigma}$, $n_i=n_{i\uparrow}+n_{\downarrow}$,
$t>0$ is the hopping parameter, $U>0$ is the Coulomb repulsion and $\mu$ is the chemical potential.  The
sums are over all sites of a three-dimensional cubic lattice, and
$\langle i,j\rangle$ denotes the sum over the nearest neighbors.

We represent the Fermi operators, the spin of the itinerant electrons
\be\label{QCB1b}
s^{\nu}_i=\frac 12 \sum\limits_{\sigma\sigma'}c^+_{i\sigma} \tau^{\nu}_{\sigma\sigma'}c^{\phantom +}_{i\sigma'},\ee where
$(\tau^x,\tau^y,\tau^z)$ are Pauli matrices, and the density operators $n_{i\sigma}$  in terms of the Schwinger bosons
($\varphi_{i,\sigma}, \varphi_{i,\sigma}^+$) and slave fermions
($h_i, h_i^+,d_i,d_i^+$). The Bose fields
are doublets $(\sigma=1,2)$ without charge, while fermions
are spinless with charges 1 ($d_i$) and -1 ($h_i$):
\begin{eqnarray}\label{QCB2} & & c_{i\uparrow} =
h_i^+\varphi _{i1}+ \varphi_{i2}^+ d_i, \qquad c_{i\downarrow} =
h_i^+ \varphi _{i2}- \varphi_{i1}^+ d_i, \nonumber
\\
& & n_i = 1 - h^+_i h_i +  d^+_i d_i,\quad  s^{\nu}_i=\frac 12
\sum\limits_{\sigma\sigma'} \varphi^+_{i\sigma}
{\tau}^{\nu}_{\sigma\sigma'} \varphi_{i\sigma'},\nonumber
\\& &
c_{i\uparrow }^+c_{i\uparrow }c_{i\downarrow }^+c_{i\downarrow}=d_i^+d_i \eea

\be\label{QCB2b}
\varphi_{i1}^+ \varphi_{i1}+ \varphi_{i2}^+ \varphi_{i2}+ d_i^+
d_i+h_i^+ h_i=1  \ee
To solve the constraint (Eq.\ref{QCB2b}), one makes a change of variables, introducing
Bose doublets $\zeta_{i\sigma}$ and
$\zeta^+_{i\sigma}\,$\cite{Schmeltzer}
\begin{eqnarray}\label{QCB3}
\zeta_{i\sigma} & = & \varphi_{i\sigma} \left(1-h^+_i h_i-d^+_i
d_i\right)^
{-\frac 12},\nonumber \\
\zeta^+_{i\sigma} & = & \varphi^+_{i\sigma} \left(1-h^+_i h_i-d^+_i
d_i\right)^ {-\frac 12},
\end{eqnarray}
where the new fields satisfy the constraint
$\zeta^+_{i\sigma}\zeta_{i\sigma}\,=\,1$. In terms of the new fields
the spin vectors of the itinerant electrons Eq.(\ref{QCB1b}) have the form
\be
s^{\nu}_{i}=\frac 12 \sum\limits_{\sigma\sigma'} \zeta^+_{i\sigma}
{\tau}^{\nu}_{\sigma\sigma'} \zeta_{i\sigma'} \left[1-h^+_i
h_i-d^+_i d_i\right] \label{QCB4} \ee
When, in the ground state,
the lattice site is empty, the operator identity $h^+_ih_i=1$ is
true. When the lattice site is doubly occupied, $d^+_id_i=1$. Hence,
when the lattice site is empty or doubly occupied the spin on this
site is zero. When the lattice site is neither empty nor doubly
occupied ($h^+_ih_i=d^+_id_i=0$), the spin equals $\,\,{\bf s}_{i}=1/2
{\bf n}_i,\,\,$ where the unit vector
\be\label{QCB5b}
n^{\nu}_i=\sum\limits_{\sigma\sigma'} \zeta^+_{i\sigma}
{\tau}^{\nu}_{\sigma\sigma'} \zeta_{i\sigma'}\qquad ({\bf
n}_i^2=1)\ee identifies the local orientation of the spin of the
itinerant electron.

The Hamiltonian Eq.(\ref{QCB1}), rewritten in terms of Bose fields Eq.(\ref{QCB3}) and slave fermions, adopts the form
\bea\label{QCB3a}
h  & = & -t\sum\limits_{\langle ij \rangle} \left[\left ( d^+_j d_i-h^+_j h_i \right) \zeta^+_{i\sigma}\zeta_{j\sigma}\right. \nonumber \\
& + & \left.\left ( d^+_j h^+_i-d^+_i h^+_j\right )\left (\zeta_{i1}\zeta_{j2}-\zeta_{i2}\zeta_{j1}\right ) + h.c. \right]\nonumber \\
& \times & \left(1-h^+_i h_i-d^+_id_i\right)^{\frac 12}\left(1-h^+_j h_j-d^+_jd_j\right)^{\frac 12} \nonumber \\
& + & U \sum\limits_i d^+_id_i -\mu \sum\limits_i \left (1-h^+_ih_i+d^+_id_i\right),\eea

An important advantage of working with Schwinger bosons and slave fermions
is the fact that Hubbard term is in a diagonal form. The fermion-fermion and fermion-boson interactions are included in the hopping term. One treats them as a perturbation. To proceed we approximate the hopping term of the Hamiltonian Eq.(\ref{QCB3a}) setting  $\left(1-h^+_i h_i-d^+_id_i\right)^{\frac 12}\sim 1$ and keeping only the quadratic, with respect to fermions, terms. This means that the averaging in the subspace of the fermions is performed in one fermion-loop approximation. Further, we represent the resulting Hamiltonian as a sum of two terms
\be\label{QCB4a}
h=h_0 + h_{int}, \ee
where
\bea\label{QCB4b}
h_0 = & - & t\sum\limits_{\langle ij \rangle} \left ( d^+_j d_i-h^+_j h_i + h.c.\right)
 +  U \sum\limits_i d^+_id_i \nonumber \\
& - & \mu \sum\limits_i \left (1-h^+_ih_i+d^+_id_i\right),\eea
is the Hamiltonian of the free $d$ and $h$ fermions, and
\bea\label{QCB4c}
h_{int} = & - &t\sum\limits_{\langle ij \rangle} \left[\left ( d^+_j d_i-h^+_j h_i \right) \left (\zeta^+_{i\sigma}\zeta_{j\sigma}-1\right)\right. \\
& + & \left.\left ( d^+_j h^+_i-d^+_i h^+_j\right )\left (\zeta_{i1}\zeta_{j2}-\zeta_{i2}\zeta_{j1}\right ) + h.c. \right]\nonumber\eea
is the Hamiltonian of boson-fermion interaction.

The ground state of the system, without accounting for the spin fluctuations, is determined by the free-fermion Hamiltonian $h_0$ and is labeled by the density of electrons
\be\label{QCB4d} n=1-<h^+_i h_i>+<d^+_id_i> \ee (see equation (\ref{QCB2})) and the "effective spin" of the electron
\begin{equation}
m=\frac 12 \left(1-<h^+_i h_i>-<d^+_id_i>\right). \label{QCB5}
\end{equation}
At half-filling

\be\label{QCB4e} <h^+_i h_i>=<d^+_id_i>. \ee To solve this equation, for all values of the parameters $U$ and $t$, one sets the chemical potential $\mu=U/2$. Utilizing this representation of $\mu$ we calculate the effective spin "m" as a function of the ratio $t/U$. The result is depicted in figure (\ref{m and J}).

Let us introduce the vector,
\begin{equation}
 M^{\nu}_{i}= m \sum\limits_{\sigma\sigma'} \zeta^+_{i\sigma}
{\tau}^{\nu}_{\sigma\sigma'} \zeta_{i\sigma'}\quad {\bf M}_{i}^2=m^2 .
\label{QCB7}
\end{equation}
Then, the spin-vector of itinerant electrons Eq.(\ref{QCB4}) can be written in the
form
\be\label{QCB5a} {\bf s}_{i}=\frac {1}{2m}{\bf M}_{i}\left(1-h^+_i\,h_i\,-\,
d^+_i\,d_i\right),\ee
where the vector  ${\bf M}_i$ identifies the local orientation of the spin of
the itinerant electrons.
The contribution of itinerant electrons to the total magnetization
is  $<{\bf s}^z_{i}>$. Accounting for the definition of $m$ (see
Eq.\ref{QCB5}), one obtains $<{\bf s}^z_{i}>= <{\bf M}^z_{i}>$.

 The Hamiltonian is quadratic with
respect to the fermions $d_i, d^+_i$ and $h_i, h^+_i$, and one can
average in the subspace of these fermions (to integrate them out in
the path integral approach). As a result, one obtains an effective
model for vectors  ${\bf M}_i$, which identifies the local orientation of the spin of the itinerant electrons,
with Hamiltonian
\begin{equation}
 h_{eff}= J \sum\limits_{  \langle  ij  \rangle  } {{\bf M}_i
\cdot {\bf M}_j}\label{QCB9}
\end{equation}
The effective exchange constant $J$ is calculated in the one
loop approximation and in the limit when the frequency and the wave
vector are small. At zero temperature, one obtains
\bea\label{QCB10} & & J = \\
& & -\frac {t}{6m^2}\frac 1N
\sum\limits_{k}\left(\sum\limits_{\nu=1}^3\cos
k_{\nu}\right)\left[\theta(-\varepsilon^d_k)-\theta(-\varepsilon^h_k)\right]
\nonumber \\
& + & \frac {2t^2}{3m^2 U}\frac 1N
\sum\limits_{k}\left(\sum\limits_{\nu=1}^3\sin^2
k_{\nu}\right)\left[1-\theta(-\varepsilon^h_k)-\theta(-\varepsilon^d_k)\right]\nonumber\eea
where $N$ is the number of lattice sites, $\varepsilon^h_k$ and
$\varepsilon^d_k$  are fermions' dispersions,
\bea\label{QCB11}
\varepsilon^h_k & = & 2t(\cos k_x+\cos k_y+\cos k_z) +\mu  \\
\varepsilon^d_k & = & -2t(\cos k_x+\cos k_y+\cos k_z)+U -\mu,
\nonumber \eea
 and the wave vector $k$ runs over the first Brillouin zone of a cubic lattice.

 The exchange constant $J$ and the effective spin $m$ are functions of the ratio $t/U$. At half-filling the exchange constant $J$ is positive and the  model (\ref{QCB9}) is an effective  model of itinerant antiferromagnetism.
 The functions $m(t/U)$ and $J(t/U)/U$ are depicted in figure (\ref{m and J}). At half-filling the density of doubly occupied states $<d^+d>$ is equal to the density of empty states $<h^+h>$. Increasing the ratio $t/U$ increases the density of doubly occupied states which in turn decreases the effective spin of the system (see equation (\ref{QCB5})).
\begin{figure}[!ht]
\centerline{\psfig{file=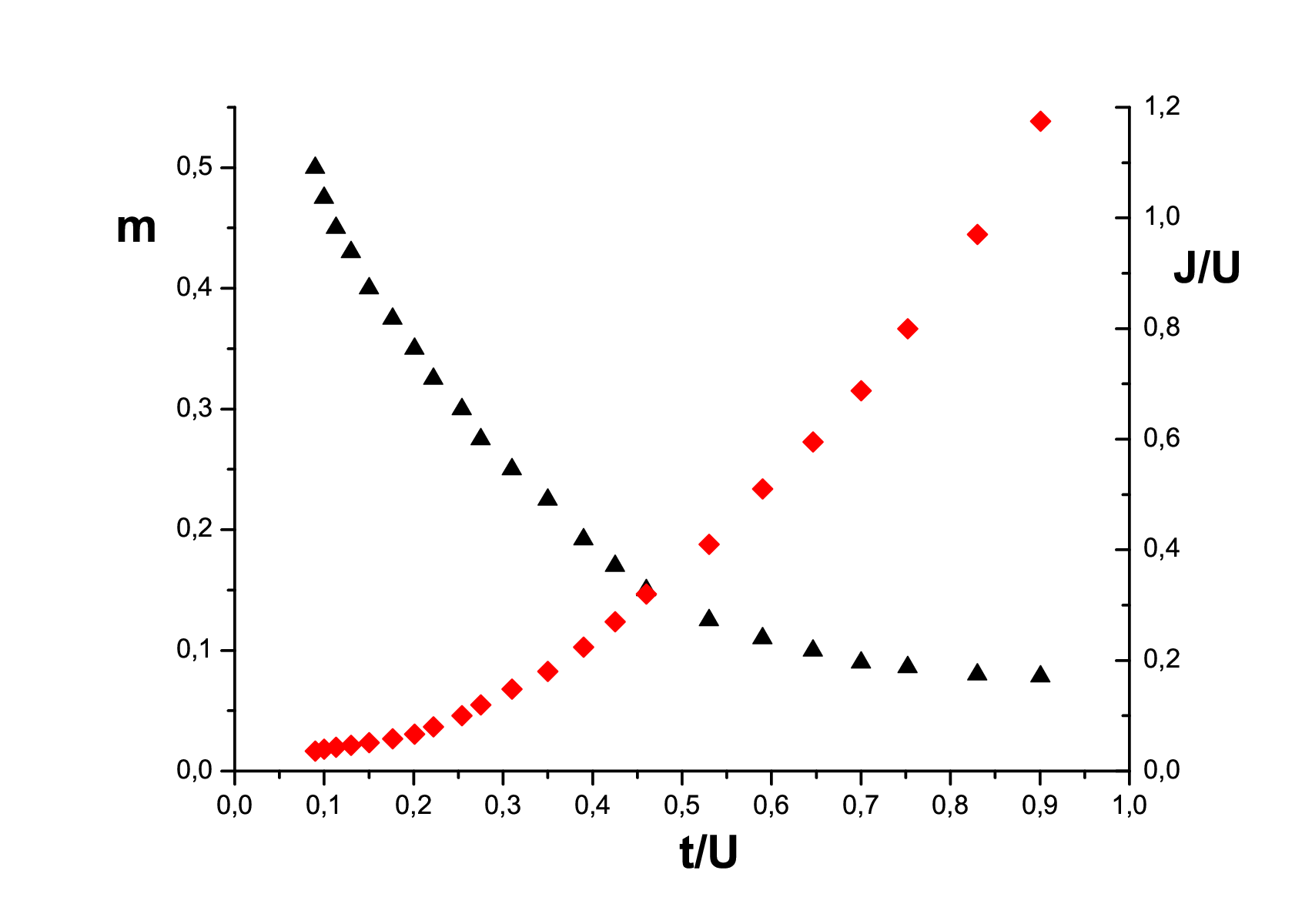,width=10.5cm,height=7cm}}
\caption{(Color online) The effective spin of the system $m$ as a function of the ratio $t/U$-black triangles(left scale). Dimensionless exchange constant $J/U$  as a function of $t/U$-red rhombuses(right scale).}
\label{m and J}
\end{figure}

\section{\bf Phase diagram}

We are going to study the antiferromagnetic phase of the model
Eq.(\ref{QCB9}) with $J>0$. To proceed, one uses the Holstein-Primakoff representation of the spin vectors ${\bf M}_j(a^+_j,\,a_j)$, where $a^+_j,\,a_j$ are Bose fields:
\begin{eqnarray}\label{QCB14}
M_j^+ & = & M_j^1+iM_j^2 \nonumber \\
& = & \cos^2\frac {\theta_j}{2} \sqrt{2m-a_j^+a_j}\, a_j - \sin^2\frac {\theta_j}{2} a_j^+ \sqrt{2m-a_j^+a_j}\nonumber \\
M_j^- & = & M_j^1-iM_j^2  \\
& = & \cos^2\frac {\theta_j}{2} a_j^+ \sqrt{2m-a_j^+a_j} - \sin^2\frac {\theta_i}{2} \sqrt{2m-a_j^+a_j}\,  a_j \nonumber \\
M_j^3 & = & \cos\theta_j (m-a_j^+ a_j) \nonumber
\end{eqnarray}
where $\theta_j={\bf Q}\cdot {\bf r}_j$ and ${\bf Q}=(\pi,\pi,\pi)$ is the antiferromagnetic wave vector.
In terms of the Bose fields and
keeping only the quadratic and quartic terms, the effective Hamiltonian
(Eq.\ref{QCB9}) adopts the form
\be\label{QCB15}
h_{eff}=h_2+h_4 \ee
where
\be\label{QCB16}
h_2\,=\,J m \sum\limits_{\langle  ij  \rangle} \left ( a_i^+a_i\, +\,a_j^+a_j\,-\,a_i^+a_j^+\,-\,a_ia_j \right )\ee
\bea\label{QCB17}
h_4 & = & \frac J4 \sum\limits_{\langle  ij  \rangle}\left ( a_i^+a_j^+a_j^+a_j\,+\,a_i^+a_i^+a_j^+a_i \right. \\
 & + & \left. a_i^+a_ia_ia_j\,+\,a_j^+a_ja_ja_i\,-\,4 a_i^+a_j^+a_ia_j \right )\nonumber \eea
and the terms without Bose fields are dropped.

The next step is to represent the Hamiltonian in the Hartree-Fock  approximation. To this end one represents the product of
two Bose fields in the form
\be\label{QCB18}
a^+_i a_j\,=\,a^+_i a_j\,-\,<a^+_i a_j>\,+\,<a^+_i a_j> \ee
and neglects all terms $(a^+_i a_j\,-\,<a^+_i a_j>)^2$ in the four magnon interaction Hamiltonian. The result is
\bea\label{QCB19}
a_i^+a_j^+a_j^+a_j & \approx & -<a_i^+a_j^+><a_j^+a_j> \nonumber \\
& +& a_i^+a_j^+<a_j^+a_j>+a_j^+a_j<a_i^+a_j^+> \nonumber \\
a_i^+a_i^+a_j^+a_i & \approx & -<a_i^+a_i><a_i^+a_j^+> \nonumber \\
& + & a_i^+a_j^+<a_i^+a_i>+a_i^+a_i<a_i^+a_j^+> \nonumber \\
a_i^+a_ia_ia_j & \approx & - <a_i^+a_i><a_ia_j>  \nonumber \\
& + & a_ia_j<a_i^+a_i>+a_i^+a_i<a_ia_j> \nonumber \\
a_j^+a_ja_ja_i & \approx & - <a_j^+a_j><a_ja_i>  \\
& + & a_ja_i<a_j^+a_j>+a_j^+a_j<a_ja_i> \nonumber \\
2a_i^+a_j^+a_ia_j & \approx & -<a_i^+a_j^+><a_ia_j> \nonumber \\ & - & <a_i^+a_i><a_j^+a_j> \nonumber \\
& + & a_i^+a_j^+<a_ia_j>+a_ia_j<a_i^+a_j^+> \nonumber \\
& + & a_i^+a_i<a_j^+a_j>+a_j^+a_j<a_i^+a_i> \nonumber \eea
We assume that the matrix elements do not depend on the lattice's links and $<a_i^+a_j^+>= <a_ia_j>$.
Then the Hartree-Fock approximation for the effective Hamiltonian (Eq.\ref{QCB15}) can be represented as a
sum
\be\label{QCB20}
h_{eff}\approx h_{HF}=h_{cl}+h_q \ee
where
\be\label{QCB21}
h_{cl}\,=\,6Jm^2N(r-1)^2 ,\ee
\be\label{QCB22}
h_q\,=\,J m r \sum\limits_{\langle  ij  \rangle} \left ( a_i^+a_i\, +\,a_j^+a_j\,-\,a_i^+a_j^+\,-\,a_ia_j \right )\ee
and $r$ is the Hartree-Fock parameter, to be determined self-consistently from the equation
\bea\label{QCB23}
r & = & 1-\frac {1}{2m}\frac 1N \sum\limits_{k} <a_k^+a_k> \\
& + & \frac {1}{2m}\frac 1N \sum\limits_{k} \frac {\cos k_x+\cos k_y+\cos k_z}{3}<a_k^+a_{-k}^+> .\nonumber \eea
Equation (\ref{QCB22}) shows that the Hartree-Fock parameter $r$ renormalizes
the exchange constant $J$.

It is convenient to rewrite the Hamiltonian (Eq.\ref{QCB22}) in momentum space representation:
\be\label{QCB24}
h_q=\sum\limits_{k}\left [\varepsilon a_k^+a_k\,-\,\gamma_k \left (a_k^+a_{-k}^+\,+\,a_ka_{-k}\right ) \right ] \ee
where
\bea\label{QCB25}
& & \varepsilon\,=\,6Jmr \\
& & \gamma_k\,=\, Jmr \left (\cos k_x + \cos k_y +\cos k_z \right ) \nonumber \eea
To diagonalize the Hamiltonian one introduces new Bose field
$\alpha_k,\,\alpha_k^+$ by means of the
transformation
\be \label{rsw12a}
a_k\,=u_k\,\alpha_k\,+\,v_k\,\alpha^+_{-k}\qquad
a_k^+\,=u_k\,\alpha_k^+\,+\,v_k\,\alpha_{-k}
\ee
where the coefficients of the transformation $u_k$ and $v_k$ are real functions of the wave vector $k$
\bea \label{QCB28}
& & u_k\,=\,\sqrt{\frac 12\,\left (\frac
{\varepsilon}{\sqrt{\varepsilon^2-4\gamma^2_k}}\,+\,1\right
)} \\
& & v_k\,=\,sign (\gamma_k)\,\sqrt{\frac 12\,\left (\frac
{\varepsilon}{\sqrt{\varepsilon^2-4\gamma^2_k}}\,-\,1\right
)}\nonumber \eea
The transformed Hamiltonian adopts the form \be
\label{QCB29} h_q = \sum\limits_{k}\left
(E_k\,\alpha_k^+\alpha_k\,+\,E^0_k\right),
\ee
with dispersion
\be\label{QCB30}
E_k\,=\,\sqrt{\varepsilon^2\,-\,4\gamma^2_k}\ee
and vacuum energy
\be\label{QCB31}
 E^{0}_k\,=\,\frac
12\,\left [\sqrt{\varepsilon^2\,-\,4\gamma^2_k}\,-\,\varepsilon \right]\ee
For positive values of the Hartree-Fock parameter and all values of $k\in B$,\,
the dispersion is nonnegative $ E_k\geq 0$. It is equal to zero at $\textbf{k}=(0,0,0)$ and
$\textbf{k}^*=(\pm\pi,\pm\pi,\pm\pi)$. Therefore, $\alpha_k$-boson describes the two long-range excitations (magnons) in the spin system \cite{note}.
Near these vectors the dispersion adopts the form $E_k\propto c_s |\textbf{k}|$ and $E_k\propto c_s |\textbf{k}-\textbf{k}^*|$
with spin-wave velocity $c_s=2\sqrt{3}Jmr$.

One can rewrite the equation for the Hartree-Fock parameter (Eq.\ref{QCB23}) in terms of the $\alpha_k$ field
\bea\label{QCB32}
 r(T) & = & 1+\frac {1}{4m}-\frac {1}{12m}\frac 1N \sum\limits_{k}\sqrt{9-e_k^2}\,\,[1+2n_k(T)]\nonumber \\
 e_k & = & \cos k_x+\cos k_y+\cos k_z\eea
where  $n_k$ is the Bose function of the $\alpha$ excitations
\be\label{QCB33}
n_k(T)=\frac {1}{e^{\frac {E_k}{T}}-1}. \ee
It is important to stress that  Eq. (\ref{QCB32}) can be obtained from the equation
\be\label{QCB34}  \partial\mathcal{F}/\partial r=0 \ee
where $\mathcal{F}$ is
the free energy of a system with Hamiltonian $h_{HF}$  (Eq.\ref{QCB20})
\be\label{QCB35} \mathcal{F} =  6J m^2 (r-1)^2+\frac 1N \sum\limits_{k}E^{0}_k
 +  \frac {T}{N} \sum\limits_{k} \ln\left(1-e^{-\frac {E_k}{T}}\right).\ee

The sublattice magnetizations $M^A$ and $M^B$ for sublattice A ($\cos \theta_i=1$) and sublattice B ($\cos \theta_i=-1$)
are defined by Eq. (\ref{QCB14}). It is evident that $M^A=-M^B$, so that the total magnetization is zero.
In terms of the Bose function $n_k$ of the $\alpha$ excitations they adopt the form
\bea\label{QCB36}
M^A(T)& = & -M^B(T)=m-\frac 1N \sum\limits_{k}<a_k^+a_k>  \\
& = & m+\frac 12-\frac 12 \frac 1N \sum\limits_{k}\frac {3}{\sqrt{9-e_k^2}}[1+2n_k(T)]
\nonumber \eea
At N\'{e}el temperature $T_N$ the sublattice magnetization is zero $M^A(T_N)=-M^B(T_N)=0$.
From equation (\ref{QCB36}) and equation (\ref{QCB32}) rewritten at the N\'{e}el temperature
one obtains a system of equations which determines the N\'{e}el temperature
\bea\label{QCB37}
r(T_N) & = & 1+\frac {1}{4m}-\frac {1}{12m}\frac 1N \sum\limits_{k}\sqrt{9-e_k^2}\,\,[1+2n_k(T_N)]\nonumber\\
2m+1 & = & \frac 1N \sum\limits_{k}\frac {3}{\sqrt{9-e_k^2}}\,[1+2n_k(T_N)]
\eea

To clarify the importance of the notion effective spin $m$ one investigates the relationship between N\'{e}el temperature and $m$.
The dependence of the dimensionless temperature $T_N/J$ on effective spin is depicted in figure (\ref{TN-m}).
\begin{figure}[!ht]
\centerline{\psfig{file=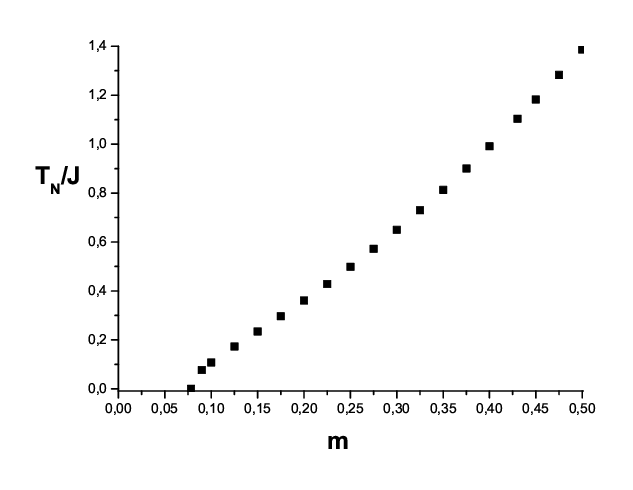,width=9cm,height=7cm}}
\caption{(Color online)\,The dependence of the dimensionless temperature $T_N/J$ on the effective spin of the itinerant electron $m$. The quantum critical value of the effective spin, the value at which $T_N=0$, is $m_{cr}=0.078$ }
\label{TN-m}
\end{figure}
Decreasing the effective spin decreases the N\'{e}el temperature. The quantum critical value of the effective spin, the value at which $T_N=0$, is $m_{cr}=0.078$. The effective spin decreases because the density of the doubly occupied states increases. The quantum critical point is a state with domination of the doubly occupied sites.

Utilizing the dependence of the effective spin $m$ and the exchange constant $J/U$ on the parameter $t/U$ (see figure \ref{m and J}) one can obtain the dependence of the dimensionless temperature $T_N/U$ on the ratio $t/U$. The phase diagram in the plane of temperature $T_N/U$ and control parameter $t/U$ is depicted in figure (\ref{TN-tU}). The quantum critical value of the ratio is $t/U=0.9$.

To compare with experimental temperature-pressure curves one has to establish the relationship between hopping parameter $t$ and pressure and between Coulomb repulsion $U$ and pressure. The simplest assumption that $U$ is a constant and $t$ is a linear function of the pressure leads to a result which well reproduces the temperature-pressure phase diagram of $CeRhIn_5$. But, to obtain the experimental phase diagrams of $CePd_2Si_2$ or $CeIn_3$ one has to implement a much more complicated fitting procedure.

\begin{figure}[!ht]
\centerline{\psfig{file=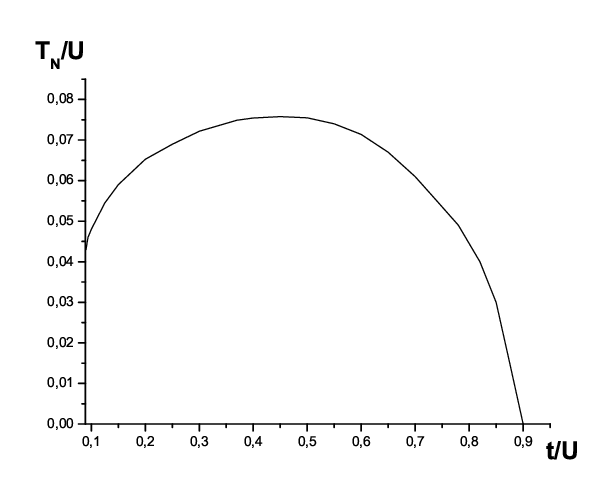,width=9cm,height=7cm}}
\caption{(Color online)\,Phase diagram in plane of temperature $T_N/U$ and control parameter $t/U$. The quantum critical value of the ratio is $t/U=0.9$}
\label{TN-tU}
\end{figure}

\section{\bf Paramagnetic phase}

When the system undergoes a thermal ($T_N>0$) or quantum ($T_N=0$) transition to a paramagnetic state, the magnon's dispersion opens a gap. This is a generic feature of the second order phase transition. To describe it mathematically, one utilizes the modified spin-wave theory proposed to describe 2D ferromagnetic \cite{Takahashi86,Takahashi87} and antiferromagnetic \cite{Takahashi89,Hirsch89} systems at finite temperature. Takahashi's idea is to supplement the spin-wave theory with the constraint that the magnetization be zero. In the present paper we formulate, along the same line, a modified spin-wave theory of the paramagnetic phase.

To enforce the magnetization on the two sublattices to be equal to zero in the paramagnetic phase,
 one introduces the parameter $\lambda$, and the new Hamiltonian  is obtained from the old one (Eq. \ref{QCB9}) by adding a new term
\be\label{QCB38} \hat{h}\,=\,h_{eff}\,-\,\lambda \sum\limits_{i}(m-a_i^+a_i)\ee

This modification leads to a modification of the Hamiltonian (Eq.\ref{QCB24}). One obtains
\be\label{QCB39}
\hat {h}_q=\sum\limits_{k}\left [\hat {\varepsilon} a_k^+a_k\,-\,\gamma_k \left (a_k^+a_{-k}^+\,+\,a_ka_{-k}\right ) \right ] \ee
with
\be\label{QCB40}
\hat {\varepsilon}= 6Jmr+\lambda \ee
We implement the same calculations as above and arrive at a Hamiltonian which is modification of the Hamiltonian (Eq.\ref{QCB29})
\be\label{QCB41}\hat {h}_q = \sum\limits_{k\in B}\left
(\hat {E}_k\,\alpha_k^+\alpha_k\,+\,\hat {E}^0_k\right),
\ee
with new dispersion
\be\label{QCB42}
\hat {E}_k\,=\,\sqrt{\hat {\varepsilon}^2\,-\,4\gamma^2_k}\ee
and new vacuum energy
\be\label{QCB43}
 \hat {E}^{0}_k\,=\,\frac
12\,\left [\sqrt{\hat {\varepsilon}^2\,-\,4\gamma^2_k}\,-\,\hat {\varepsilon} \right]\ee
The free energy $\hat {\mathcal{F}}$ of a system with the modified Hamiltonian reads
 \be\label{QCB44} \hat {\mathcal{F}} =  6J m^2 (r-1)^2+\frac 1N \sum\limits_{k}\hat {E}^{0}_k
 +  \frac {T}{N} \sum\limits_{k} \ln\left(1-e^{-\frac {\hat {E}_k}{T}}\right).\ee
Then, one can obtain the system of equations for the Hartree-Fock parameter and the parameter $\lambda$ from the equations
\be\label{QCB45}  \partial\mathcal{F}/\partial r=0,\qquad \partial\mathcal{F}/\partial {\lambda}=0. \ee
The result is
\bea\label{QCB46}
r(T) & = & 1+\frac {1}{4m} \\
& - & \frac {1}{12m}\frac 1N \sum\limits_{k}\frac {3\hat {\varepsilon}-2Jmr e_k^2}{\sqrt{\hat {\varepsilon}^2-4\gamma_k^2}}[1+2\hat {n}_k(T)]\nonumber\\
2m+1 & = & \frac 1N \sum\limits_{k}\frac {\hat {\varepsilon}}{\sqrt{\hat {\varepsilon}^2-4\gamma_k^2}}[1+2\hat {n}_k(T)]\nonumber
\eea
where $\hat {n}_k$ is the Bose function of $\alpha$ excitation (Eq.\ref{QCB33}) with new dispersion $\hat {E}_k$ (Eq.\ref{QCB42}).

It is convenient to represent the parameter $\lambda$ in the form
\be \label{ferri32}
\lambda\,=\,6 J m u \kappa \ee
introducing a new parameter $\kappa$.
Near to the zero wave vector the dispersion adopts the form $\hat {E}_k \propto\sqrt{c^2_s|\textbf{k}|^2+36Jmr(2\kappa+\kappa^2)}$, where
$36Jmr(2\kappa+\kappa^2)$ is the gap of the magnon. It is zero below the N\'{e}el temperature and increases when the temperature increases above the N\'{e}el temperature.

We implement the following procedure to calculate  the Hartree-Fock parameter $r$ and the parameter $\kappa$. At temperatures below the N\'{e}el temperature $\kappa=0$ and $r(T/J)$ is obtained from equation (\ref{QCB32}). At temperatures above the N\'{e}el temperature the functions $r(T/J)$ and $\kappa(T/J)$ are solutions of the system (\ref{QCB46}). The result is depicted in figures (\ref{HF}) and (\ref{kappa}).

 Fig. (\ref{HF}) shows that the renormalization $r$ at zero temperature, due to the magnon-magnon interaction, is stronger when the system approaches the quantum critical point (curve "a"), and it is insignificant for spin-localized systems (curve "d").
\begin{figure}[!ht]
\centerline{\psfig{file=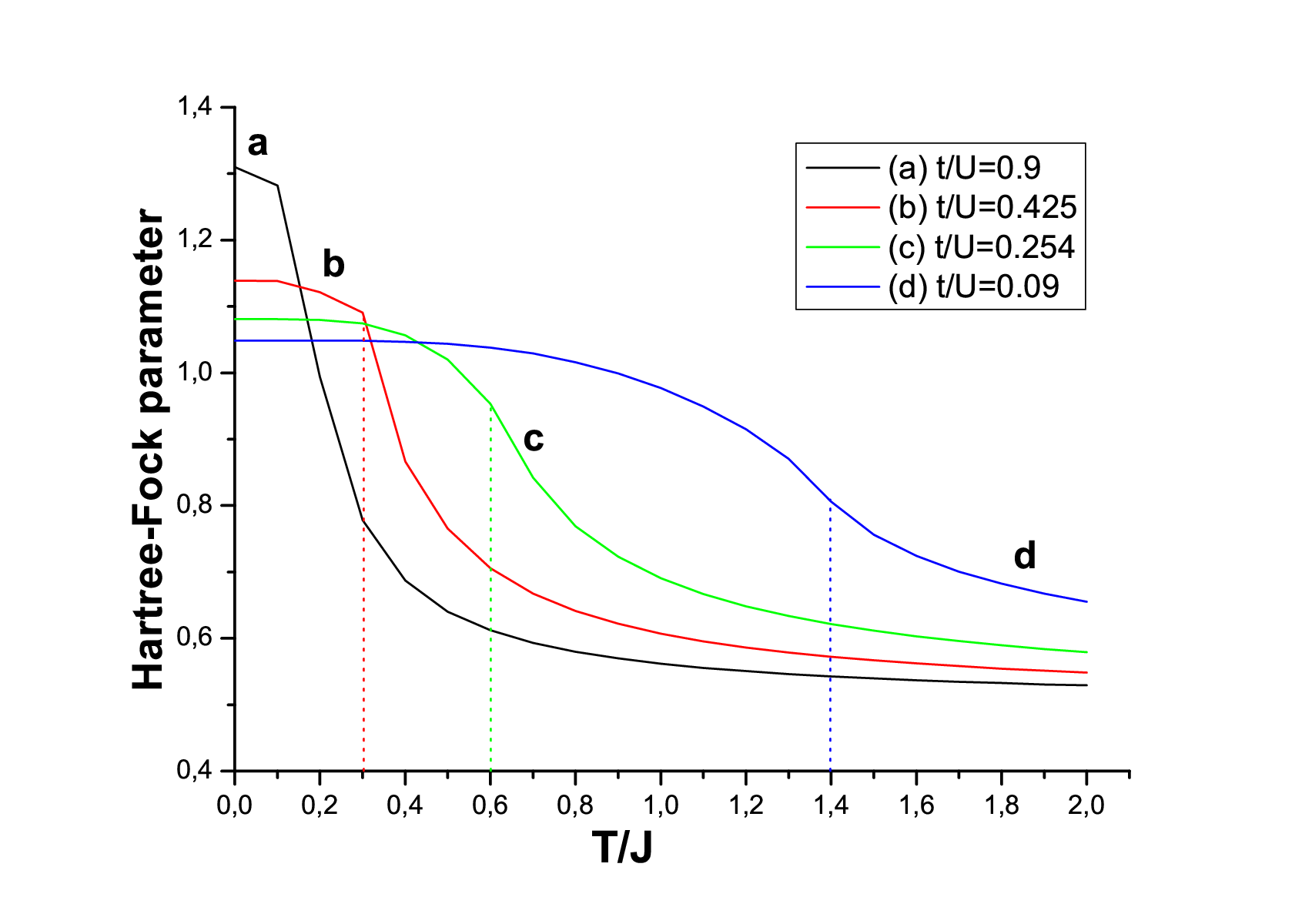,width=11cm,height=7cm}}
\caption{(Color online)The dependence of the Hartree-Fock parameter on the dimensionless temperature $T/J$: (a) at quantum critical point $t/U=0.9\,\, (m_{cr}=0.078)$; (b) at $t/U= 0.425\,\, (m=0.175)$; (c) at $t/U=0.254\,\, (m=0.3)$; (d) at $t/U=0.09\,\, (m=0.5)$. The vertical dotted lines correspond to the N\'{e}el temperatures $T_N/J$.}
\label{HF}
\end{figure}

The $\kappa$ parameter is a measure for the gap of the magnon in the paramagnetic phase. It is zero at the N\'{e}el temperature and increases when the temperature increases. The function $\kappa(T/J)$ is depicted in figure (\ref{kappa}) for different values of the ratio $t/U$.
\begin{figure}[!ht]
\centerline{\psfig{file=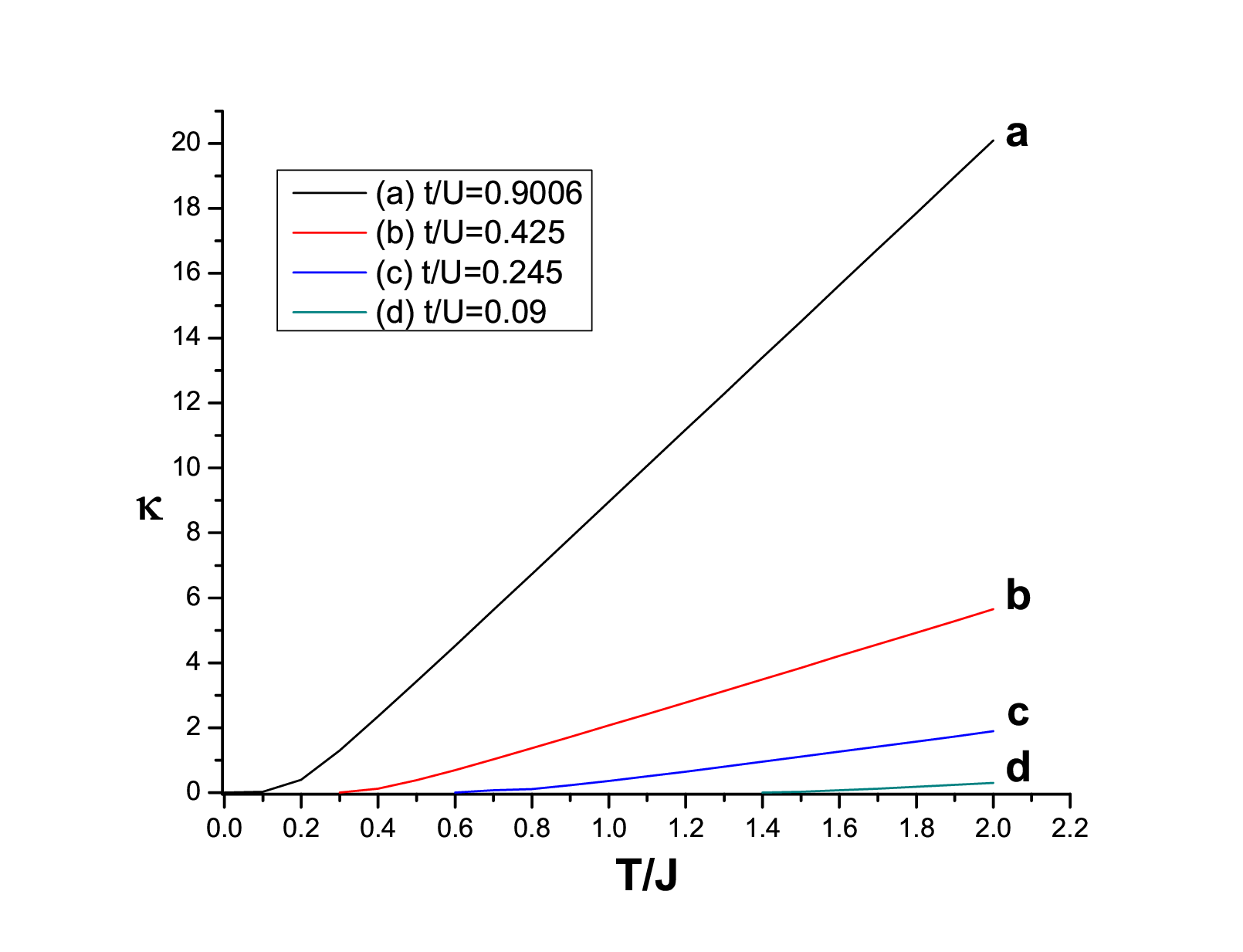,width=11cm,height=7cm}}
\caption{(Color online) The dependence of $\kappa$ parameter on the dimensionless temperature $T/J$: (a) at quantum critical point $t/U=0.9$; (b) at $t/U= 0.425$; (c) at $t/U=0.254$; (d) at $t/U=0.09$.}
\label{kappa}
\end{figure}

Fig.(\ref{kappa}) shows that the increase is faster when the system approaches the quantum critical point (curve "a"), and it is weak for spin-localized systems (curve "d").

\section{\bf Specific heat}

Utilizing the above calculated functions $r(T/J)$ and $\kappa(T/J)$ one can calculate the magnons' contribution to the specific heat of the system. By definition, the entropy is
\be\label{QCB48}   \mathcal{S}=-\frac {d\mathcal{F}}{dT}=-\frac {\partial\mathcal{F}}{\partial r}\frac {\partial r}{\partial T}-
\frac {\partial\mathcal{F}}{\partial \lambda}\frac {\partial \lambda}{\partial T}-\frac {\partial\mathcal{F}}{\partial T}\ee
where $\mathcal{F}$ is the free energy of the system Eqs.(\ref{QCB35},\ref{QCB44}). Owing to Eqs. (\ref{QCB34}) and (\ref{QCB45}) the first two terms in Eq.(\ref{QCB48}) are equal to zero and one obtains the customary formula for the entropy of a Bose system
\be\label{QCB49}
\mathcal{S}=\frac 1N \sum\limits_{k}\left [(1+n_k)\ln (1+n_k)-n_k\ln n_k \right ], \ee
where the dispersion $E_k$ Eq.(\ref{QCB30}) is used to define the Bose function Eq.(\ref{QCB33}) below the N\'{e}el temperature, and dispersion $\hat {E_k}$ Eq.(\ref{QCB42}) above it.
With entropy, as a function of temperature in mind, one can calculate the contribution of magnons to the specific heat:
\be\label{QCB50}
C_v=T \frac {d \mathcal{S}}{dT} \ee
The resultant curves $C_v(T/J)$, for different values of the parameter $t/U$, are depicted in figure (\ref{Cv}).
\begin{figure}[!ht]
\centerline{\psfig{file=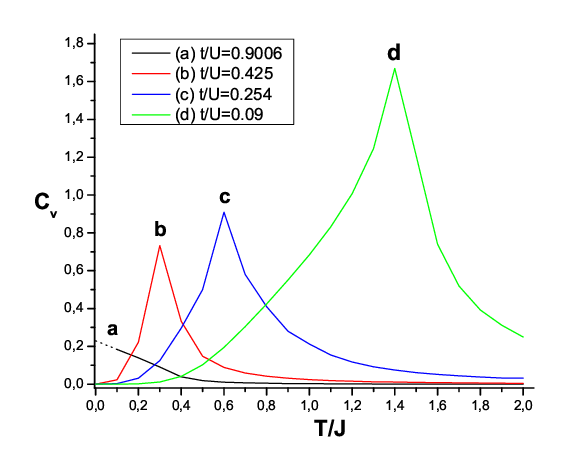,width=11cm,height=7cm}}
\caption{(Color online)Contribution of magnons' fluctuations to the specific heat: (a) at quantum critical point $t/U=0.9\,\, (m_{cr}=0.078$; (b) at $t/U= 0.425\,\, (m=0.175)$; (c) at $t/U=0.254\,\, (m=0.3)$; (d) at $t/U=0.09\,\, (m=0.5)$.}
\label{Cv}
\end{figure}

The function $C_v(T/J)$ has a maximum at the N\'{e}el temperature. Fig. (\ref{Cv}) shows that the maximum is suppressed at the quantum critical point (curve "a"). The maximum is well observed experimentally\cite{Hegger00,Knebel04}, and one can use it to determine the N\'{e}el temperature. The experimental measurements of the specific heat of $CeRhIn_5$ for different pressures \cite{Knebel04} show that at $T_N$, $C_v/T$ has a very sharp peak at ambient pressure. With increasing pressure the magnetic anomaly remains well defined but the maximum decreases. This phenomenon is well described theoretically in the present paper Fig.(\ref{Cv}).
The existing correspondence between specific heat and the derivative of the resistivity suggests a confidence that magnon's fluctuations are important and for the transport properties of the itinerant antiferromagnets  .

\section{\bf Summary}

In this paper itinerant antiferromagnets were studied. Varying the ratio $t/U$, where $t$ is the hopping parameter and $U$ is the Coulomb repulsion, the system was investigated between a state with localized spins ($t/U<0.09$) and a quantum critical point $t/U=0.9$ at which the N\'{e}el temperature is zero. The evolution of the magnons' fluctuations in antiferromagnetic and paramagnetic phases was studied by means of the renormalized spin-wave theory and modified spin-wave theory. The renormalized spin-wave theory includes a parameter $r$ which accounts for magnon-magnon interactions. The system of equations for the N\'{e}el temperature (\ref{QCB37}) rewritten for a 2D system has the only solution $T_N=0$. This shows that the present method of calculation is in accordance with the Mermin-Wagner theorem, which is our theoretical criterion for an adequate account of magnons' fluctuations. The modified spin-wave theory involves in a natural way the gap of the magnon in the paramagnetic phase.

The effective spin is the most important factor which determines the quantum criticality. To figure this out one has to map the low-energy excitations of the half-filled Hubbard model onto an effective spin-1/2 Heisenberg model \cite{Szczech95}. The dimensionless  N\'{e}el temperature $T_N/J$ for
this model is $T_N/J=1.387$. The result shows that the  N\'{e}el temperature of the spin-1/2 antiferromagnets is nonzero for all values of the exchange constant including the weak coupling $U/t<<1$ regime \cite{Szczech95}. The mapping of the Hubbard model onto the spin-1/2 Heisenberg model correctly describes the magnetism of localized electrons ($t/U<0.09$). This is the so-called Heisenberg limit. If we want to account for the process of  delocalization in the system we have to introduce the notion of the effective spin of the electron. The delocalization, in a half-filled system, is accompanied by increasing the density of doubly occupied states which in turn decreases the effective spin and pushes the system to the quantum critical point. Introducing this notion one can map the Hubbard model onto  the Heisenberg like model but with effective spin smaller than the spin (1/2) of the electron. This permits us to go beyond the Heisenberg limit accounting for the process of delocalization and at the same time to use the same technique of calculation as in the case of the Heisenberg model of localized spins.
The mapping of the Hubbard model onto an effective Heisenberg model with effective spin $m<1/2$ is a crucial step towards the understanding of quantum criticality.

The critical value of the effective spin $m=0.078$ only depends on the effective Heisenberg model.
The effective exchange constant $J$ and the effective spin $m$ are calculated in one fermion-loop approximation. It is neither a strong coupling approximation nor a weak coupling approximation. At half-filling the chemical potential is fixed $\mu=U/2$ and the dispersions' Eq.(\ref{QCB11}) dependence on the parameter $t/U$ is nontrivial.

To compare with experimental results the phase diagram in the plane of $T/U$ and control parameter $t/U$ was obtained. To compare with results of the numerical calculations one has to convert the diagram Fig.(\ref{TN-tU}) into the phase diagram in the plane of $T/t$ and $U/t$. The result is shown in figure  (\ref{TN-Ut}).
\begin{center}
\begin{figure}[htb]
\centerline{\psfig{file=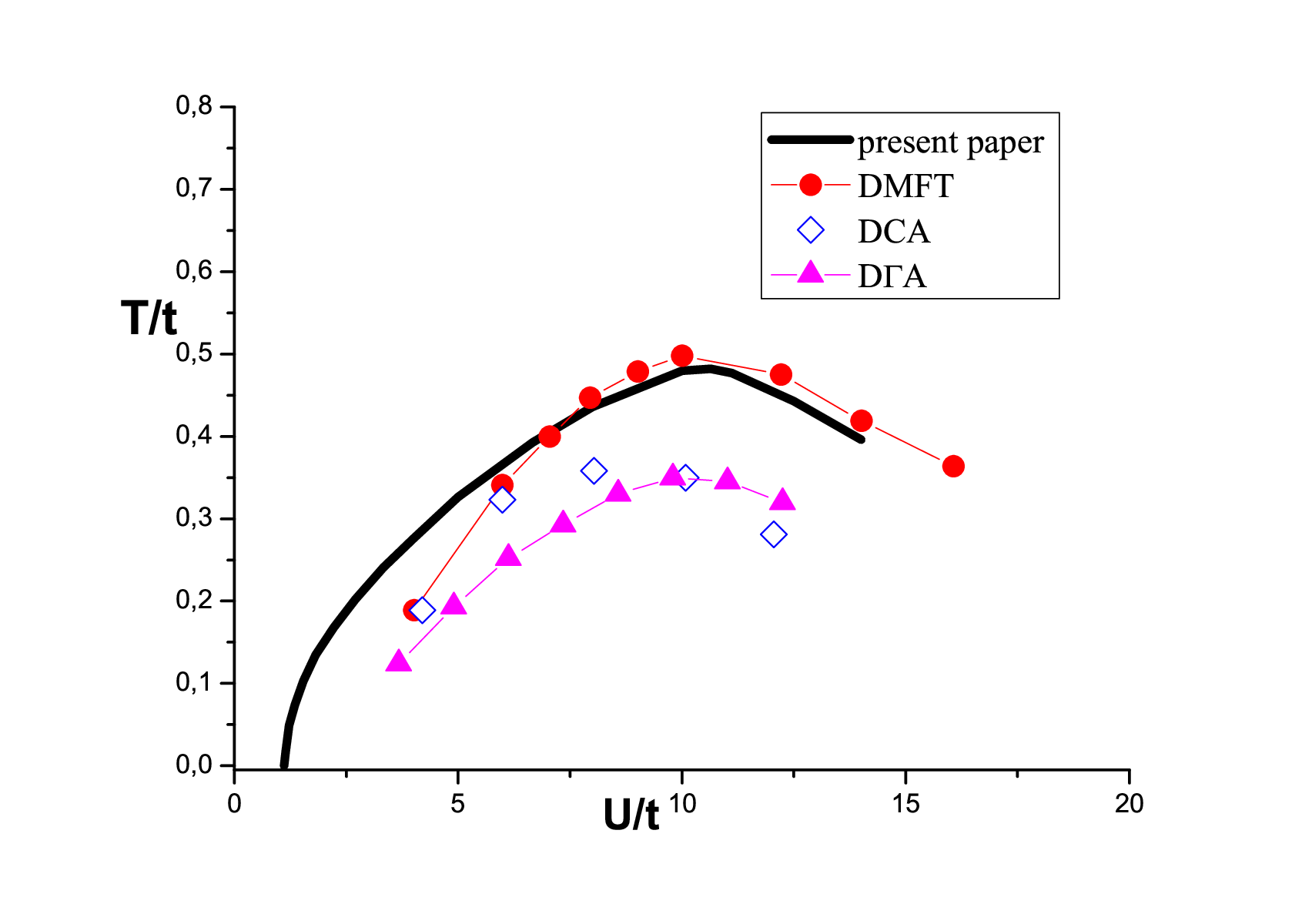,width=11cm,height=7cm}}
\caption{(color online)\,Phase diagram in plane of temperature $T_N/t$ and control parameter $U/t$. The solid black line is the phase diagram obtained  in the present paper. The red circles, the blue open diamonds and the magenta triangles are the results in (DMFT)\cite{Kent05}, (DCA)\cite{Kent05} and (D$\Gamma$A)\cite{Toschi07} respectively.}
\label{TN-Ut}
\end{figure}
\end{center}
The solid black line is the phase diagram obtained in the present paper. The red circles, the blue open diamonds and the magenta triangles are the results obtained by means of the dynamical mean field theory (DMFT) \cite{Kent05},  dynamical cluster approximation (DCA)\cite{Kent05} and dynamical vertex approximation (D$\Gamma$A) \cite{Toschi07}, respectively.

One of the most important points in the Hubbard model is the maximum of the  N\'{e}el temperature $T_N/t$ as a function of $U/t$. It indicates the crossover from itinerant magnetism to the magnetism of localized spins. The N\'{e}el temperature  of itinerant systems increases, when $U/t$ increases, as a result of the increasing of the effective spin. The N\'{e}el temperature of the spin-1/2 Heisenberg model of localized electrons is $T_N=1.387 J$ with an exchange constant which decreases when $U/t$ increases. In the present paper the maximum is at $U/t=10.6$, in the DMFT  at $U/t=10$ and in D$\Gamma$A at $U/t=9.8$. The overall agreement of the results for the  N\'{e}el temperature in the present paper with the results obtained in DMFT are satisfactory for all values of $U/t$ except for the lowest one. The non-local corrections, accounted for in D$\Gamma$A, reduce the  N\'{e}el temperature $T_N/t$ versus temperatures in DMFT and the present paper in the whole phase diagram.

The exchange constant $J$ of the effective model Eq. (\ref{QCB9}) is calculated in the limit when the frequency and the wave vector are small.
The calculations can be improved employing an effective Heisenberg theory with exchange constant $J(k)$ which depends on the wave vector $k$.

Another important point in the Hubbard model is the quantum critical point (QCP) $T_N=0$. The phase diagram Fig.(\ref{TN-Ut}) shows that it is reached at $U/t=1.11$. The figure also shows that all numerical calculations are implemented at control parameters $U/t>3.674$, which is far from the QCP. This explains the absence of comments on the quantum criticality. On the other hand an investigation of the Hubbard model without explicit affirmation of the existence or nonexistence of QCP in the Hubbard model is not complete. One can extrapolate the curves in DMFT, DCA and D$\Gamma$A down to zero temperature. A nonzero QCP emerges in these approaches.

Alternatively, the Hubbard model is studied by a mapping on a spin-1/2 generalized Heisenberg model with higher order terms in a form of long-range or ring exchange \cite{Reischl04}. An  effective Heisenberg  model with spin-1/2 is considered which cannot describe the quantum criticality despite the fact that parameter $t/U$ is close to the quantum critical point. This is because the quantum critical behavior depends decisively on the effective spin of the itinerant electron which is smaller than 1/2 near the quantum critical point (see figure (\ref{TN-m})).

Ne\'{e}el order and the quantum critical point in the Hubbard model are also studied by means of a spin-charge rotating reference frame approach \cite{Zaleski08}, which explicitly factorizes the charge and spin contribution to the electron operator. The effective constants are calculated by means of the Hartree-Fock approximation of the fermion interaction. This explains the over-estimation of the critical value of the control parameter $U/t=0.676 (t/U=1.479)$ compared with the result in the present paper $t/U=0.9$, where the Coulomb repulsion is treated exactly.

\section{Acknowledgments}

This work was partly supported by a Grant-in-Aid DO02-264/18.12.08 from the NSF-Bulgaria.


\begin{thebibliography}{99}
\bibitem{note} The vertexces of the Brillouin zon are topologically equivalent. This is why we have two long-range excitations
in the theory.
\end{thebibliography}
\end{document}